\documentclass[10pt,conference]{IEEEtran}
\usepackage{graphicx}
\usepackage{amsmath,amssymb}
\usepackage{booktabs}
\usepackage{makecell}
\usepackage{caption}
\usepackage{pgfplots}
\usepackage{url}
\usepackage{filecontents}
\usepackage{tcolorbox}
\usepackage{cite}
\usepackage{multirow}
\usepackage{colortbl} 
\usepackage{xcolor} 
\usepackage{subcaption}
\usepackage[colorlinks=true, linkcolor=black, citecolor=blue, urlcolor=blue]{hyperref}

\title{Improving LLM-Based Fault Localization with External Memory and Project Context}
\author{
\IEEEauthorblockN{Inseok Yeo}
\IEEEauthorblockA{
\textit{School of Computing} \\
\textit{KAIST} \\
Daejeon, Republic of Korea \\
yinseok38@kaist.ac.kr}
\and
\IEEEauthorblockN{Duksan Ryu}
\IEEEauthorblockA{
\textit{Department of Software Engineering} \\
\textit{Jeonbuk National University} \\
Jeonju, Republic of Korea \\
duksan.ryu@jbnu.ac.kr}
\and
\IEEEauthorblockN{Jongmoon Baik}
\IEEEauthorblockA{
\textit{School of Computing} \\
\textit{KAIST} \\
Daejeon, Republic of Korea \\
jbaik@kaist.ac.kr}
}

\newcounter{romantable}

\begin{document}
\maketitle

\begin{abstract}
Fault localization, the process of identifying software components responsible for failure, is important yet time-consuming. Recent advances in Large Language Models (LLMs) have made it possible to perform fault localization without defect datasets or fine-tuning. However, existing LLM-based fault localization techniques rely solely on the generic capabilities of the LLM without incorporating project-specific knowledge, such as the project's high-level purpose or unique debugging and failure patterns specific to target project, limiting their effectiveness. As a result, these methods depend on complex workflow to analyze the project from scratch, and they also perform poorly on complex software projects that require domain-specific knowledge for successful debugging.

To overcome these limitations, we propose MemFL, a novel LLM-based fault localization technique that incorporates project-specific knowledge through external memory knowledge editing. The external memory consists of two parts: static memory, which holds pre-generated summaries of the project and its classes, and dynamic memory, which iteratively gather debugging guidance from prior attempts. These memory components enhance the LLM's ability to reason effectively about complex, project-specific contexts, enabling MemFL to structure the debugging process into three simple steps that significantly reduce debugging costs while enhancing accuracy. Furthermore, the iterative refinement of dynamic memory allows the LLM  to refine its reasoning based on previous fault localization attempts progressively.

In our evaluation on the Defects4J benchmark, MemFL using GPT-4o-mini localized 12.7\% more buggy methods than existing LLM-based baselines. Remarkably, it achieved this improvement with only 21\% of the execution time (17.4 seconds per bug) and 33\% of the API cost (\$0.0033 per bug). The performance gain was even more pronounced on complex projects where LLM-based baselines struggled, localizing 27.6\% more buggy methods. Furthermore, MemFL equipped with GPT-4.1-mini significantly outperformed all existing baselines by 24.4\%, requiring just 24.7 seconds and \$0.0094 per bug. These results highlight MemFL’s core contribution: its external-memory design effectively incorporates project-specific knowledge into the LLM, greatly improving accuracy while maintaining speed and cost efficiency.

\end{abstract}

\section{Introduction}
Fault Localization (FL), the process of pinpointing the buggy software component, remains a significant bottleneck in software development. Studies indicate that developers dedicate approximately 66\% of their debugging time to fault isolation \cite{bohme2017bug}. To alleviate this, researchers have developed various automated fault localization techniques. Among the most prominent traditional techniques are spectrum-based fault localization (SBFL) \cite{de2016spectrum}, mutation-based fault localization (MBFL) \cite{6823877}, and learning-based fault localization (LBFL) \cite{10562102}.

SBFL techniques leverage program spectra by comparing the frequency of software component execution in failed versus passed test cases, statistically ranking them based on their likelihood of containing faults \cite{de2016spectrum}. In contrast, MBFL introduces random code changes (mutants) and evaluates their impact on test outcomes to localize fault more precisely. However, both techniques heavily depend on a comprehensive test suite that thoroughly executes various execution paths. MBFL, in particular, is costly due to mutation, which further limits its practicality \cite{10.1016/j.jss.2009.06.035}\cite{unknown}.

In contrast, LBFL techniques, such as GRACE \cite{10.1145/3468264.3468580} and DeepFL \cite{10.1145/3293882.3330574}, employ machine-learning models to learn complex patterns from the target project, demonstrating superior performance. However, training LBFL models requires large, high-quality defect datasets and extensive training. In practice, applying machine-learning to real-world software projects often suffers from data scarcity, leading to suboptimal model performance \cite{Alzubaidi2023ASO,whang2023data}. This issue is severe in fault localization, where defect datasets are especially difficult to obtain and often exhibit challenges such as mislabeled data, noise, and class imbalance \cite{8502824,croft2023data}.

Recent advancements in Large Language Models (LLMs) provide a promising solution to these limitations. LLMs, pre-trained on vast and diverse datasets, can be applied to various downstream tasks without requiring task-specific training data\cite{minaee2025largelanguagemodelssurvey}. Different LLM-based FL techniques, such as LLMAO\cite{yang2024large} and Wu et al.'s approach\cite{wu2023large}, were proposed. However, these techniques were limited to file-wide or class-wide input, making it impractical for project-wide fault localization, limiting the practicality.

Several LLM-based fault localization techniques have been proposed to handle the challenge of debugging large codebases with a limited input size. SoapFL \cite{10891926} addresses this limitation through a multi-agent approach, where specialized LLM-based modules collaborate and divide the fault localization task to enable project-wide fault localization. Similarly, AutoFL \cite{10.1145/3660771} uses an LLM tool-use to navigate software repositories autonomously rather than processing the entire repository as a single input, allowing project-wide fault localization.

These techniques effectively perform fault localization using readily available inputs such as source code, error messages, test cases, method documentation, and static analysis result. However, they overlook LLM's potential for understanding deeper project-specific context, including the high-level purpose of the project, source code context, and unique debugging and failure patterns. Prior research indicates that incorporating project-specific knowledge can significantly enhance bug comprehension and fault localization \cite{8918999,10.1007/s10664-022-10190-x}. However, current LLM-based fault localization methods rely solely on the model’s built-in knowledge, neglecting this additional context. This leads to two critical limitations: First, they depend heavily on complex multi-agent systems or intricate LLM tool usage to process complex software projects from scratch. Second, their effectiveness significantly declines on more complex projects requiring detailed project-specific knowledge. These could limit the practicality and scalability of current LLM-based fault localization approaches.

To address these limitations, we introduce MemFL, a novel fault localization approach that integrates project-specific knowledge with external memory knowledge editing \cite{10.1145/3698590}. Instead of fine-tuning the LLM,  we employ external memory-based knowledge editing by simply concatenating an external memory containing project-specific knowledge to deliver the necessary information. The external memory that contains project-specific information guides the model to localize fault through enhanced reasoning with a simple three-step process: Bug Review Generation, Code Condensation, and Fault Confirmation, resulting in improved performance and reduced cost. Moreover, we propose a two-component external memory structure that efficiently enhances performance by incorporating debugging guidance acquired from previous fault localization attempts. The first component, static memory, contains pre-generated project information that remains constant throughout debugging, providing information related to existing classes and target projects. The second component, dynamic memory, is iteratively updated based on insights gained during previous debugging attempts. By selectively refining the dynamic memory to incorporate valuable debugging guidance acquired from prior fault localization, MemFL further improves its performance cost-effectively.

We tested MemFL on the Defects4J \cite{10.1145/2610384.2628055} benchmark, which includes 350 real Java bugs, and compared it with leading LLM-based methods (SoapFL, AutoFL) and other baselines (GRACE, DeepFL, Ochiai). Using GPT-4o-mini, MemFL showed about 12.7\% higher Top-1 accuracy than other LLM-based techniques, while keeping costs much lower. On complex projects where previous LLM-based methods struggled, MemFL achieved an even greater improvement, 27.6\% higher Top-1 accuracy. With the latest GPT-4.1-mini model, MemFL outperformed all baselines by a large margin. Furthremore, we analyzed optimal debugging guidance generation policy, and showed effectiveness of each component of MemFL through ablation study. These results show that MemFL’s simple three-step approach, using project-specific knowledge from external memory, is both effective and efficient for fault localization.

The main contributions of this paper are:
\begin{itemize}
\item We propose MemFL, a novel LLM-based fault localization approach that leverages project-specific knowledge, enabling effective and efficient fault localization even on projects requiring complex domain-specific understanding.
\item We introduce a two-component external memory structure that facilitates efficient delivery and refinement of project-specific knowledge while maintaining computational efficiency.
\item We present an optimal dynamic memory generation policy and identify the best model architecture through ablation studies.
\item We evaluate MemFL on the Defects4J benchmark and demonstrate that it outperforms existing baselines.
\end{itemize}

\stepcounter{romantable}

\section{Background and Related Works}
\label{sec:related_works}
This section provides a detailed overview of existing fault localization techniques, categorizing them into Spectrum-Based Fault Localization (SBFL), Mutation-Based Fault Localization (MBFL),  Learning-Based Fault Localization (LBFL), and LLM-based Fault localization. Additionally, this section introduces background information about knowledge editing.

\subsection{Spectrum-Based Fault Localization (SBFL) and Mutation-Based Fault Localization (MBFL)}
\label{subsec:sbfl_mbfl}
SBFL \cite{10.1016/j.jss.2009.06.035,10.5555/1308173.1308264,10.1109/PRDC.2006.18,5431781,1007991,10.1145/2000791.2000795,6651713,4291037,10.1007/978-3-642-33119-0_18} leverages program spectra, or the history of different program executions, to pinpoint faulty program components. SBFL compares execution frequencies in failing versus passing test cases, assigning suspiciousness scores to rank software components by their likelihood of containing a fault. MBFL \cite{6823877,10.1002/stvr.1509,10.1145/3551349.3556949,10.1145/3377812.3382146}, on the other hand, employs program mutations to identify fault locations more precisely. MBFL introduces small changes (mutations) into the code and observes the impact on test outcomes, more accurately inferring the location of faults.

While SBFL and MBFL are effective in various scenarios, they share limitations. Both techniques heavily depend on a comprehensive test suite that thoroughly explores various execution paths \cite{10.1016/j.jss.2009.06.035, unknown}. Furthermore, MBFL is particularly resource-intensive due to the mutation and testing required \cite{unknown}.

\subsection{Learning-Based Fault Localization (LBFL)}
\label{subsec:lbfl}

LBFL techniques have emerged as a promising alternative, leveraging machine-learning models to overcome the limitations of SBFL and MBFL. LBFL techniques train models on defect datasets to learn complex patterns that correlate code characteristics and test execution data with fault locations \cite{10562102}. This allows LBFL techniques to capture intricate relationships within the software project and test outcomes, leading to improved performance \cite{10.1145/3468264.3468580}.

GRACE \cite{10.1145/3468264.3468580} utilizes machine-learning to predict the probability of a faulty code component. GRACE extracts various code features, such as code complexity metrics and change history, and trains a machine-learning model to perform fault localization. By learning from labeled defect datasets, GRACE can more effectively localize faults. On the other hand, DeepFL employs deep learning models to enhance fault localization accuracy \cite{10.1145/3293882.3330574}. DeepFL leverages deep neural networks to learn more complex and abstract representations of code and execution data compared to traditional machine-learning models. 

Despite their advancements, LBFL techniques face challenges. The most significant one is their dependency on large, labeled datasets for training. Acquiring and labeling such datasets is often computationally expensive and impractical in real-world contexts \cite{Alzubaidi2023ASO,whang2023data}. This is especially true for defect datasets, which often suffer from mislabeled examples, noisy annotations, and severe class imbalance \cite{8502824,croft2023data}. Furthermore, LBFL models often operate as black boxes, listing faulty code without providing debugging steps \cite{10.1145/3660771,10891926}, which can reduce developer trust and hinder debugging \cite{10.1007/978-3-031-65392-6_23}.

\begin{table}[h]
    \centering
    \captionsetup{labelformat=empty} 
    \caption{\textbf{TABLE \Roman{romantable}:} Code metrics for projects in Defects4j} \label{tab1}
    \begin{tabular}{l|cccc}
        \toprule
        Project & KLoC & TestKLoC & Tests & \makecell{Avg. \# of Classes \\ Covered by Bugs} \\
        \midrule
        Chart   & \textbf{96}  & 50  & 2205  & 26.6  \\
        \textbf{Closure} & \textbf{90} & \textbf{83}  & \textbf{7927} &\textbf{78.6}  \\
        Math    & 85  & 19  & 3602  & 9.0   \\
        Time    & 28  & 53  & 4130  & 48.3  \\
        Lang    & 22  & 6   & 2245  & 2.21  \\
        \bottomrule
    \end{tabular}
\end{table}

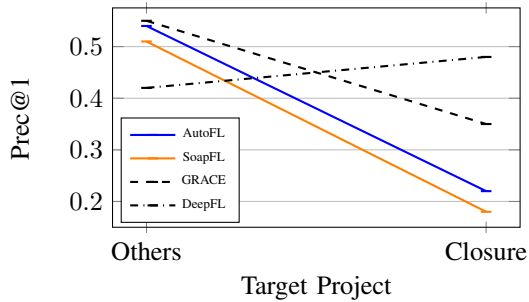
\begin{figure}
    \centering
    \begin{tikzpicture}
        \begin{axis}[
            width=7cm, height=4.5cm, 
            xlabel={Target Project},
            ylabel={Prec@1},
            symbolic x coords={Others, Closure},
            xtick=data,
            ymin=0.15, ymax=0.575,
            ymajorgrids=true,
            xtick align=center, 
            legend style={
                at={(0.02,0.02)}, 
                anchor=south west,
                font=\tiny 
            },
            mark options={solid},
            xticklabel style={rotate=0}, 
            yticklabel style={/pgf/number format/.cd, fixed, precision=2} 
        ]

        \addplot[color=blue, thick, mark=-] coordinates {
            (Others,0.54)
            (Closure,0.22)
        };
        \addlegendentry{AutoFL}

        \addplot[color=orange, thick, mark=-] coordinates {
            (Others,0.51)
            (Closure,0.18)
        };
        \addlegendentry{SoapFL}

        \addplot[color=black, dashed, thick, mark=-] coordinates {
            (Others,0.55)
            (Closure,0.35)
        };
        \addlegendentry{GRACE}

        \addplot[color=black, dash dot, thick, mark=-] coordinates {
            (Others,0.42)
            (Closure,0.48)
        };
        \addlegendentry{DeepFL}

        \end{axis}
    \end{tikzpicture}

    \caption{FL performance comparison in Closure and other projects between LLM-based baseline techniques and LBFL}
    \label{fig:fl_comparison}
\end{figure}

\subsection{LLM-based Fault Localization}
\label{subsec:llm_based_fl}
Recent advancements in Large Language Models (LLMs) offer promising new opportunities for fault localization by addressing key limitations of traditional techniques. LLMs, pre-trained on vast amounts of code and natural language data, possess an inherent understanding of programming languages, code structures, and bug patterns \cite{zhao2025surveylargelanguagemodels}. Extensive pre-training on large datasets equips LLMs to perform a wide array of downstream tasks without fine-tuning \cite{minaee2025largelanguagemodelssurvey}. This characteristic overcomes a major limitation of LBFL: the need for large, high-quality defect datasets and fine-tuning. 

Wu et al. \cite{wu2023large} investigated ChatGPT's effectiveness for fault localization, examining how prompt length and context size affect performance. They found ChatGPT promising, especially at the function-wide input, but noted reduced accuracy with larger code contexts. LLMAO \cite{yang2024large} trained a lightweight adapter on top of CodeGen \cite{nijkamp2023codegenopenlargelanguage}, enabling accurate and test-free fault localization while requiring only a small amount of data for training. However, both techniques are unsuitable for project-wide input due to input size limitations. However, prior research indicates that developers prefer project-wide fault localization with method-level granularity \cite{10.1145/2931037.2931051}, and many recent LLM-based debugging approaches are designed to operate on top of method-level fault localization \cite{10.1109/ICSE48619.2023.00125,10.1109/ICSE48619.2023.00129}. This highlights a key challenge: enabling project-wide fault localization within a limited context window of LLMs.

SoapFL \cite{10891926} introduces a multi-agent architecture for project-wide fault localization. It employs four specialized LLM-driven agents—Test Code Reviewer, Source Code Reviewer, Software Architect, and Software Test Engineer—that collaborate to analyze bug behavior, navigate complex code structures, and confirm the root cause of failures. The fault localization process in SoapFL is structured into three main steps and seven substeps, allowing the LLM to handle smaller, more manageable inputs, which enables project-wide fault localization.

AutoFL \cite{10.1145/3660771} leverages LLMs with structured tool-use to autonomously explore software repositories and localize faults. By enabling LLMs to invoke functions that retrieve method definitions, documentation, and test coverage data, AutoFL can identify the root cause. AutoFL repeats this process five times, combining its results to improve accuracy. This iterative tool-use allows the system to gather relevant information about the target software project despite the limitations of an LLM’s fixed context length, allowing project-wide fault localization.

\subsection{Knowledge Editing}
\label{subsec:Knowledge_Editing}

Knowledge editing is the process of modifying LLM to incorporate specific knowledge while minimizing the impact on unrelated pre-trained knowledge. Approaches generally fall into three categories: global modification, local modification, and external memorization \cite{10.1145/3698590}. Global methods, such as RecAdam \cite{chen2020recalllearnfinetuningdeep} and Editable Training \cite{sinitsin2020editableneuralnetworks}, adjust a large set of parameters under constraints to preserve unrelated abilities. Local methods identify and tweak only the parameters holding the target fact; examples include ROME \cite{meng2023locatingeditingfactualassociations} and prompt-based MEMIT \cite{inproceedings}. Although both can deliver additional knowledge to LLM, they often incur high computational costs, risk overfitting, and may inadvertently alter unrelated knowledge \cite{10.1145/3698590}.

External memory approaches keep new knowledge separate from the core model. Extension-based techniques—like T-Patcher \cite{huang2023transformerpatchermistakeworthneuron}, and CALINET \cite{dong2022calibratingfactualknowledgepretrained}—add small trainable modules to the network. Memory-based strategies—such as SERAC \cite{mitchell2022memorybasedmodeleditingscale} and MemPrompt \cite{madaan2023memoryassistedprompteditingimprove}—store facts in an external key–value memory or cache. The model retrieves updated facts from this memory during inference, enabling scalable, robust updates without retraining. In this work, we use external memory knowledge editing to inject project-specific knowledge without fine-tuning.

\begin{figure}[htbp]
    \centering
    \includegraphics[width=\columnwidth]{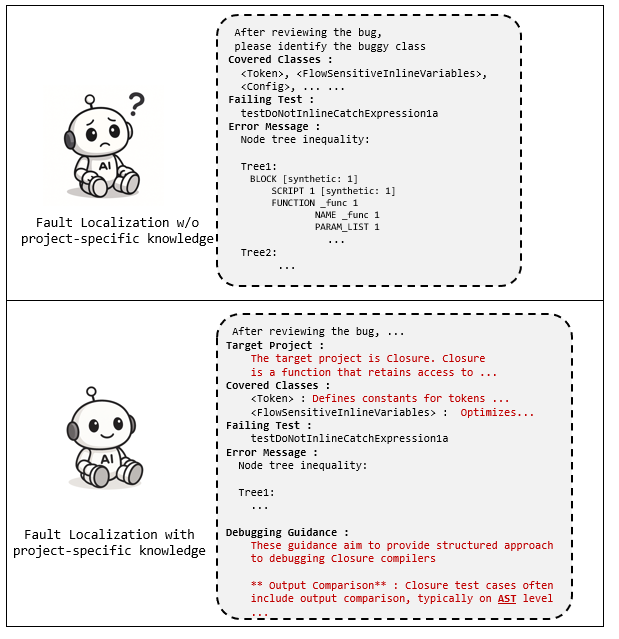 } 
    \caption{Motivating Example}
    \label{fig2}
\end{figure}

\section{Challenges and Motivation}

Although LLM-based fault localization techniques have shown promising results without requiring large error datasets, they still face notable challenges. These techniques primarily rely on readily available inputs such as source code, error messages, test cases, method documentation, and static analysis. While useful, these inputs alone do not fully utilize an LLM’s potential for understanding deeper, project-specific context. This overlooked context includes a project's overall purpose, detailed summaries of classes, and specific debugging and failure patterns unique to each project. Previous research has shown that incorporating detailed, project-specific knowledge significantly improves bug comprehension and fault localization accuracy for developers \cite{8918999,10.1007/s10664-022-10190-x}.

The absence of a deeper project-specific context introduces two major challenges. First, current LLM-based fault localization methods depend on complex multi-step workflows that involve multiple agents and extensive tool-use to analyze software projects from scratch. Without access to prior contextual knowledge, these techniques must go through a complicated reasoning, resulting in high computational overhead and prolonged execution times \cite{xia2024agentlessdemystifyingllmbasedsoftware}.

Second, performance notably declines when handling complex software projects that demand extensive contextual information for effective debugging. As illustrated in Figure 1, AutoFL and SoapFL generally perform well on most Defects4J projects except for Closure, where their effectiveness significantly drops. This decline is particularly notable compared to LBFL methods like GRACE and deepFL, which maintain relatively consistent performance across all projects.  This decline primarily arises from Closure’s complex debugging requirements, such as dealing with detailed Abstract Syntax Tree (AST) comparisons and custom assertion logic necessary for verifying optimized outputs \cite{closurecompiler}. 

Additionally, Closure stands out as the most complex project within the Defects4J dataset, as highlighted in Table 1. It has a notably higher number of classes covered per bug, expanding the search space and complicating debugging. This complexity has historically posed substantial challenges even for non-learning-based fault localization techniques, with some methods excluding Closure altogether due to these difficulties \cite{10.1145/3510003.3510073}.

Motivated by these challenges, our research introduces a simple yet effective FL technique using external memory-based knowledge editing that integrates project-specific knowledge directly into the LLM prompt. This external memory provides essential context, enabling more efficient and accurate fault localization without relying on complex processes. Additionally, the proposed two-component external memory structure supports dynamic updates, allowing iterative refinement to enhance localization performance further. Consequently, our method achieves cost-effective and precise fault localization, even in highly complex and context-rich projects.

Figure 2 demonstrates how integrating project-specific knowledge could help fault localization through simplified example from actual defects4j data (Closure, Bug 3). Without this additional context, the LLM struggles to pinpoint the fault because it lacks comprehensive awareness of the project's purpose, individual class functions, and the intricate meaning behind error messages involving complex AST comparisons. In contrast, MemFL leverages external memory containing concise project summaries, class descriptions, and targeted debugging guidance. This enables the model to correctly interpret errors, focus precisely on relevant code paths, and efficiently identify the faulty method.

\begin{figure*}[htbp]
    \centering
    \includegraphics[width=\textwidth]{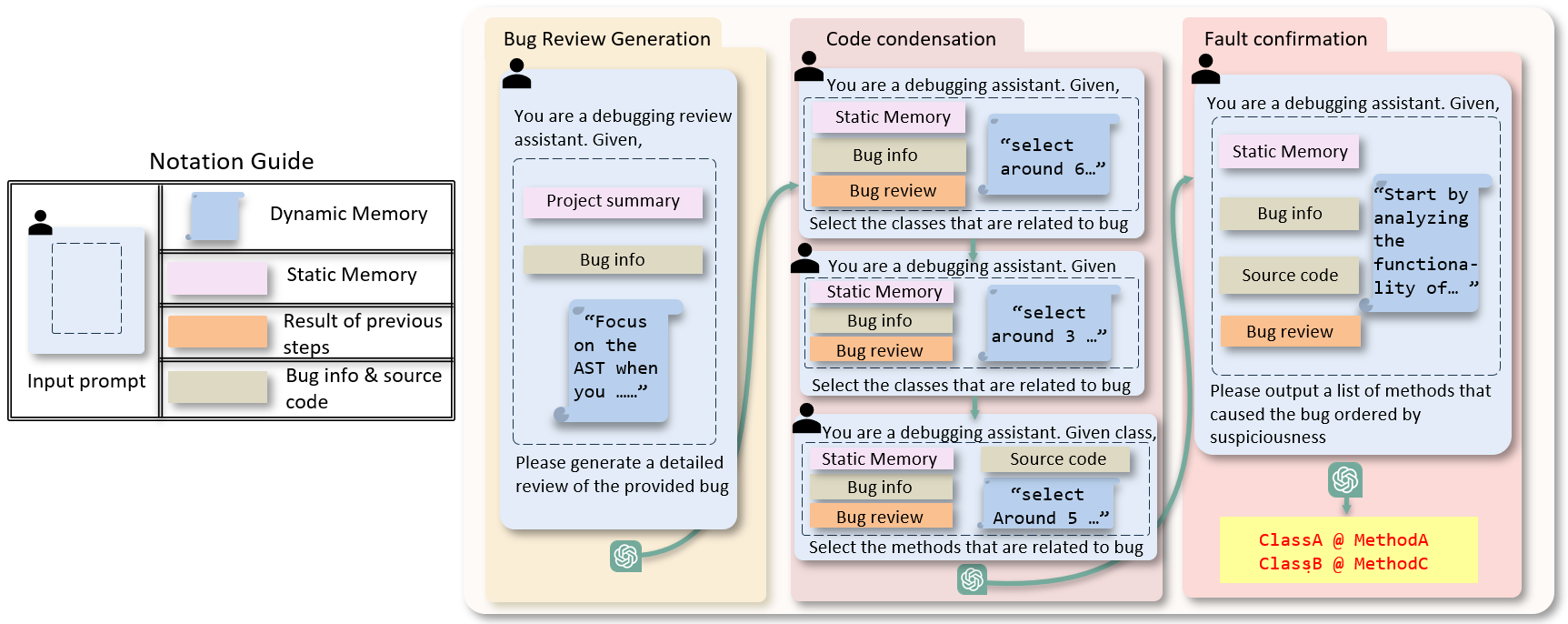} 
    \caption{Overview of Fault Localization}
    \label{fig2}
\end{figure*}

\section{Method}

In this work, we propose MemFL, a novel fault localization approach that incorporates project-specific knowledge through external memory knowledge editing. We will first detail the structure of the external memory, then provide an overview of MemFL, and finally discuss its two main stages.

\subsection{External Memory Structure}
MemFL leverages external memory to incorporate project-specific knowledge into each stage of fault localization to improve performance. This external memory is divided into two distinct types: Static Memory and Dynamic Memory. Figure 2 illustrates a simplified FL prompt that includes both types of external memory. The information highlighted in red represents the external memory. Information about the target project and covered class corresponds to Static Memory, while the debugging guidance represents Dynamic Memory.

Static Memory is generated prior to the fault localization process and remains unchanged. It consists of two parts: a project summary and a class summary. The project summary provides a high-level overview of the project's purpose and architecture, offering broader contextual information about the target project. The class summary delivers more granular information detailing the functionalities of individual classes within the project.

Dynamic Memory is adaptive and evolves throughout the fault localization. It delivers debugging guidance obtained from previous fault localization attempts, iteratively. Initially, MemFL only uses Static Memory to localize the faults. Then, the LLM reviews the results from each fault localization step, comparing debugging reports with outputs to generate and refine Dynamic Memory. This process enhances accuracy over time, ensuring the delivery of more relevant project-specific information that can be obtained through prior attempts. Importantly, we maintain separate Dynamic Memory for each fault localization step because each step requires different forms of guidance. By customizing Dynamic Memory for each specific task, MemFL provides highly relevant information that improves both accuracy and efficiency.

The two-component memory structure allows MemFL to further enhance FL performance efficiently by updating only the necessary information. A detailed overview and prompt for external memory generation will be provided in later section. 

\subsection{MemFL Framework Overview}
Figures 3 and 4 illustrate the overall structure of MemFL, which consists of two main processes: Fault Localization and External Memory Generation. The process begins with the generation of static memory and bug reports. Next, MemFL performs fault localization on a small training dataset to to build dynamic memory for each stage iteratively. This is done by comparing the outputs of each fault localization step with the corresponding bug reports to extract helpful debugging guidance. Once both external memory is prepared, the actual fault localization begins. A detailed overview of each process will be provided in a later section.

\subsection{Fault Localization}
 The Fault Localization process consists of three steps: Bug Review Generation, Code Condensation, and Fault Confirmation. Figure 3 provides a detailed overview of the Fault Localization and simplified prompt templates used in each step. This section will introduce each of these steps.

\subsubsection{Bug Review Generation}
The Bug Review Generation Step identifies the cause of a failing test case by providing the LLM with a project summary, bug details, and dynamic memory. The LLM then generates a detailed bug review describing the test's objective and the root cause of failure. This analysis ensures a deeper contextual understanding for later steps.

As shown in Figure 3, the input prompt at this step includes a project summary. This summary, part of the Static Memory added before fault localization, provides context about the project's goal. Following the project summary, bug information, including error messages, stack trace, and test codes, is appended. The error message provides the type of failure to LLM, and stack trace information consists of the sequence of method calls leading to the failure automatically generated when test failure. The test code that reported the failure is also added to the prompt. If multiple test functions report failures, only one is included to avoid exceeding the LLM's input length limitations. 

Furthermore, any other test method called by the failing test method is also included. This is crucial because, while simple assertion failures cause some bugs, many others exhibit behaviors that are more easily understood by examining the interactions between the test method. This helps provide a complete picture of the test execution and the context surrounding the failure. As a result, this step outputs a bug review that helps guide the following steps.

Dynamic Memory for the Bug Review Generation step provides insights into understanding the failure. Dynamic memory enhances the Bug Review Generation step by providing debugging guidance of typical error messages and stack trace patterns of each project. Some bugs from certain software projects involve intricate project-specific patterns, like AST comparisons or custom assertions, which are hard to interpret without context. Dynamic Memory aims to deliver this crucial information, helping the LLM analyze failures correctly,and identify root causes more precisely.

\subsubsection{Code Condensation}
Code condensation aims to reduce the amount of code LLM needs to analyze in the final step by selecting components directly related to the observed test failure. In this step, the LLM is given the following inputs: static memory, bug information, bug review, source code, and dynamic memory. The static memory includes the project summary and class summaries. The source code of selected classes is used only in the third sub-step of Code Condensation. Finally, dynamic memory provides guidance on the optimal strategy for condensing the code.

The condensation occurs in three sub-steps. The first two steps involve progressively reviewing class summaries (instead of the source code) to narrow the list of relevant classes. The third step focuses on individual classes, providing both the class review and the source code, and tasks the LLM with selecting a subset of methods for further analysis in the final Fault Confirmation step. This incremental approach helps overcome the context window size of the LLM by condensing the target project hierarchically, allowing the model to perform the final Fault Confirmation step on the most relevant code components.

Before initiating the three-step condensation process, we employ a class reduction technique inspired by \cite{10891926}. We only consider classes with a high method-level coverage rate $r_c$. This rate is calculated as: $r_c = \sum_{i=1}^{N} \mathbb{I}_{y_i} / N$, where $\mathbb{I}_{y_i}$ is a function to indicate whether method ${y_i}$ is covered, and N is the total number of methods in class C. Based on this method-level coverage rate, We select only the bugs where the faulty class ranks within the top 60. This pre-processing step successfully reduces the input size for the Code Condensation step while retaining over 98\% of the bugs.

The first two sub-steps of Code Condensation use project summaries, bug information, test review, class summaries, and dynamic memory to narrow down the classes most relevant to the bug. Only the test code that directly triggered the bug is considered to stay within the LLM's input limits. The third sub-step processes each selected class individually, using its source code and class summary to further condense it by selecting relevant methods. To avoid confusion with duplicate method names, the LLM is asked to generate methodName@lineNumber pairs. 

During the Code Condensation step, dynamic memory for each substep provides project-specific guidance on the optimal condensation strategy. It includes simple instructions on what to focus on and how many classes and methods to retain in each substep.

\subsubsection{Fault Confirmation}
Finally, Fault Confirmation is performed on the condensed code resulting from the previous step. The LLM is provided with static memory, bug information, condensed source code, bug review, and dynamic memory. The objective of this stage is for the LLM to generate a ranked list of suspicious methods, ordered by their likelihood of containing the fault.

Because methods within different classes might share the same name, we include line numbers within the source code provided to the LLM and instruct it to generate its output in the format className@methodName@lineNumber. This format ensures the unambiguous identification of each method.

The Dynamic Memory in this stage provides effective strategies for pinpointing the fault within the condensed code specific to the target project. This offers valuable insights to enhance the Fault Confirmation step.

\begin{figure*}[htbp]
    \centering
    \includegraphics[width=\textwidth]{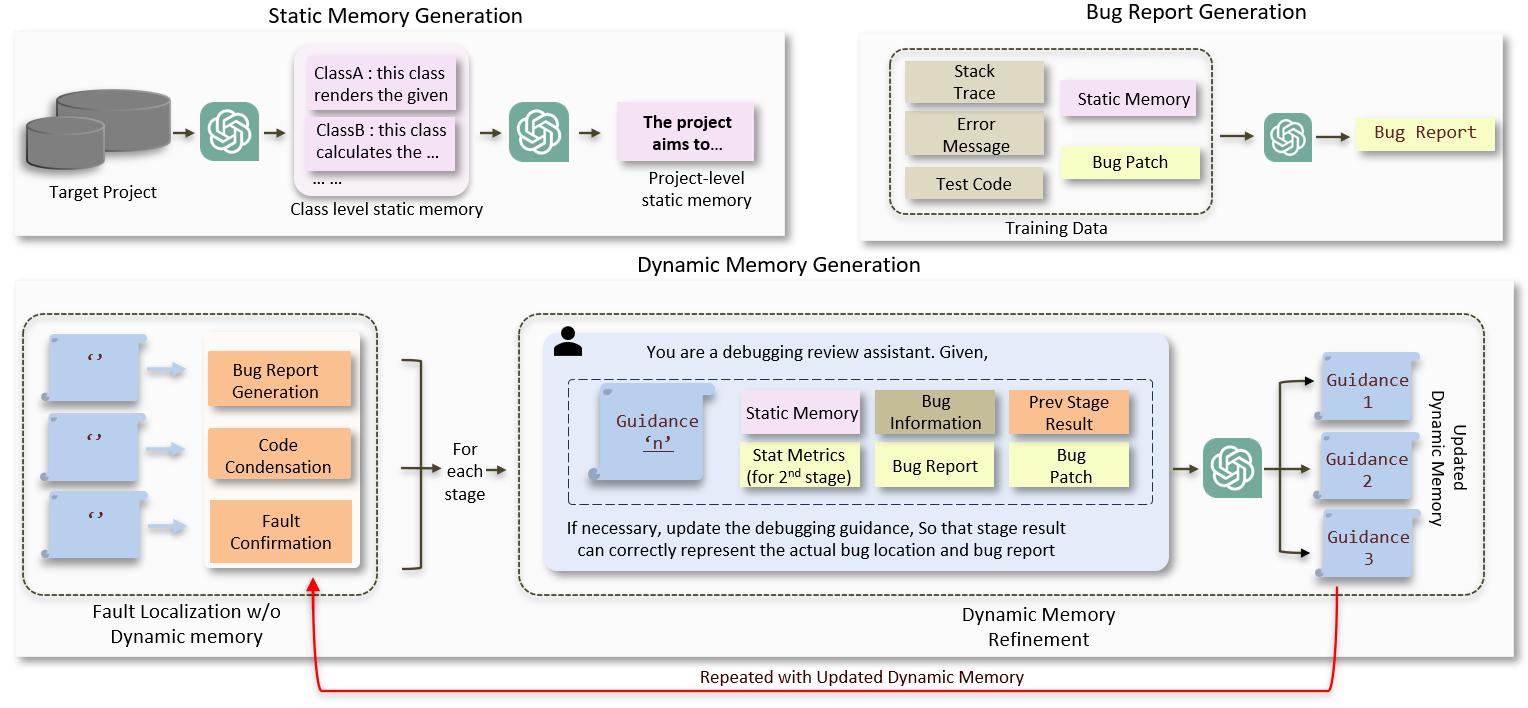} 
    \caption{Overview of External Memory Generation}
    \label{fig2}
\end{figure*}

\subsection{External Memory Generation}
External Memory Generation stage generates and refines two components of external memory: static memory and dynamic memory. Figure 4 provides a detailed overview and simplified prompt used in each stage. This section will introduce the External Memory Generation strategy of each component. 

\subsubsection{Static Memory Generation}
Static Memory provides fixed project and class-level summaries prepared before fault localization or dynamic memory generation to supply essential contextual information. To generate class summaries, we query the LLM to summarize each class based on its content. The project summary is then generated by providing the model with the project name and the aggregated class summaries, enabling a more comprehensive and context-aware description of the entire project.

\subsubsection{Dynamic Memory Generation}
Dynamic Memory is updated iteratively during fault localization to deliver refined debugging guidance based on insights from prior fault localization attempts. Dynamic Memory in MemFL is generated through a two-phase process. In the Bug Report Generation stage, a small batch of bugs is selected, and detailed bug reports are created to serve as reference cases. In the Dynamic Memory Generation stage, the LLM first runs the fault localization pipeline using only static memory; after each step, it is re-prompted with the step’s output alongside the corresponding bug report to refine the dynamic memory. The enriched dynamic memory is used in subsequent localization passes. Repeating this cycle progressively refines the guidance, enabling more accurate and context-aware fault localization.

The Bug Report Generation stage aims to collect bug reports that are essential for Dynamic Memory Generation automatically. Before entering the Bug Report Generation stage, we select a small, representative batch of bugs to serve as pseudo–training data for dynamic memory construction. For each bug, we assemble its bug information, including stack trace, error message, and failing test code, along with the source and patched versions of the buggy methods and the associated class summaries. This information is provided to the LLM, which produces a detailed, structured bug report for each case. The resulting reports are then saved and used for the Dynamic Memory generation stage.

The Dynamic Memory Generation stage iteratively produces and refines debugging guidance from earlier localization runs. It begins by running the fault localization pipeline on the training bugs using only Static Memory, then capturing the intermediate outputs at each step. Those outputs are compared with the information collected previously. For each localization phase, we prompt the LLM with (1) the intermediate result, (2) existing Dynamic Memory, and (3) the bug information, asking it to either refine the dynamic memory or confirm that no further updates are needed. When the LLM determines that refinement is needed, the Dynamic Memory is updated and incorporated into the subsequent localization pass. This cycle repeats for up to a small fixed number of iterations to progressively enhance accuracy.

The Dynamic Memory generation stage differs slightly for each dynamic memory. For dynamic memory of Bug Review Generation and Fault Confirmation, it delivers project-specific debugging guidance by re-prompting the LLM with detailed bug reports from previous iterations, ensuring maximum context extraction. In contrast, during Code Condensation, dynamic memory provides specific guidance for Code Condensation, such as the number of classes and methods to be selected. Therefore, the dynamic memory generation for the Code Condensation stage is also provided with quantitative metrics, such as the number of classes and methods selected in the last attempt, with their recall and precision. These statistical metrics help the model balance thoroughness against input size at each condensation step for each project.

The process is repeated multiple times; the optimal number of repetitions and the number of bugs used for debugging guidance generation are discussed in the results section.

\section{Experimental Design}

\subsection{Research Questions}
We investigate the following research questions:
\begin{itemize}
    \item \textbf{RQ1: What are the performance and cost of MemFL compared to baseline techniques?}
    \item \textbf{RQ2: What is the optimal dynamic memory generation for MemFL?}
    \item \textbf{RQ3: What is the impact of different design choices on the performance of MemFL?}
\end{itemize}

\subsection{Benchmark}
We evaluate MemFL using 350 active bugs from Defects4J v1 \cite{10.1145/2610384.2628055, Sobreira_2018}, a widely used benchmark of real-world bugs from five open-source Java projects. Defects4J remains a standard dataset in fault localization research, offering reproducible bugs and consistent infrastructure that support fair comparison across studies. 

\subsection{Baselines}
We compare MemFL with existing LLM-based fault localization techniques: SoapFL and AutoFL. Both techniques were originally evaluated using GPT-3.5-turbo-0613, which is now unavailable. Therefore, we re-evaluated them and MemFL using GPT-4o-mini, one of the most popular LLMs. We also evaluated MemFL using GPT-4.1-mini, a recently released model, to assess generalizability across different LLMs.  However, we encountered a reproduction issue when in the 'Math' project for SoapFL. As a result, we relied on the reported results generated using GPT-3.5-turbo-0613.

Additionally, we compare MemFL with LBFL and SBFL techniques, including GRACE, DeepFL, and Ochiai. The evaluation results were taken from SoapFL  \cite{10891926}, which modified DeepFL by excluding mutation-related features and retaining only source code and coverage-related features, referring to this modified version as DeepFL$_{\text{cov}}$
Regarding the dataset, when detailed data was available, we presented results on 350 active bugs. In cases where some data was unavailable, we report results on 357 bugs, including active and deprecated ones.

\subsection{Evaluation Metrics}
We evaluate the performance of fault localization using the top k accuracy (acc@k) metric, which measures the number of bugs accurately located within the top k suggestions. The acc@k metric (k = 1, 3, 5) is the most commonly used evaluation measure in fault localization, as prior studies \cite{10.1145/2931037.2931051} \cite{10.1145/2001420.2001445} have suggested that only the top few results are practically helpful for developers.

Moreover, we analyzed the computational cost and execution time of LLM-based fault localization techniques, measuring time in seconds and cost in terms of total API expenses in dollars. Since the evaluation on one of the five projects for SoapFL was unavailable, our primary evaluation for average cost and time for SoapFL relied on other four projects. Lastly, we conducted an overlap analysis to assess how effectively MemFL identifies unique bugs not captured by other LLM-based techniques.

We evaluated MemFL using a 5-fold cross-validation \cite{inbook} approach to prevent data leakage. Specifically, a predetermined number of data points are randomly selected from four folds (the pseudo-training set) to perform dynamic memory generation, and the resulting dynamic memory is then evaluated on the remaining fold (the test set). This procedure is repeated five times, each time using a different fold as the test set. The final performance metric is computed by summing the results obtained from these five folds. Details regarding the optimal number of selected data points and the dynamic memory generation iterations are discussed in the results section.

\section{Results}

\subsection{RQ1: Fault Localization Performance and Cost Analysis }

\begin{table}[h]
    \caption{FL Results Comparison Using LLM-Based Baselines (GPT-4o-mini)}
    \centering
    \small
    \renewcommand{\arraystretch}{1.5} 
    \setlength{\tabcolsep}{2pt} 
    \definecolor{lightgray}{gray}{0.9} 
    \begin{tabular}{l|c||ccc|ccc|ccc}
    \hline
    \multirow{2}{*}{\textbf{Project}} & \multirow{2}{*}{\footnotesize \textbf{\# Bugs}} & \multicolumn{3}{c|}{\textbf{MemFL}} & \multicolumn{3}{c|}{\textbf{AutoFL}} & \multicolumn{3}{c}{\textbf{SoapFL}} \\
    &  & {\footnotesize \textbf{Top1}} & {\footnotesize \textbf{Top3}} & {\footnotesize \textbf{Top5}} & {\footnotesize \textbf{Top1}} & {\footnotesize \textbf{Top3}} & {\footnotesize \textbf{Top5}} & {\footnotesize \textbf{Top1}} & {\footnotesize \textbf{Top3}} & {\footnotesize \textbf{Top5}} \\
    \hline\hline
    Chart & {\footnotesize 26} & {\footnotesize 14} & {\footnotesize 19} & {\footnotesize 21} & {\footnotesize 17} & {\footnotesize 23} & {\footnotesize 23} & {\footnotesize 15} & {\footnotesize 17} & {\footnotesize 17} \\
    \hline
    Lang & {\footnotesize 61} & {\footnotesize 47} & {\footnotesize 57} & {\footnotesize 58} & {\footnotesize 39} & {\footnotesize 55} & {\footnotesize 58} & {\footnotesize 35} & {\footnotesize 48} & {\footnotesize 49} \\
    \hline
    Math & {\footnotesize 106} & {\footnotesize 62} & {\footnotesize 77} & {\footnotesize 80} & {\footnotesize 60} & {\footnotesize 82} & {\footnotesize 89} & \underline{\footnotesize 57} & \underline{\footnotesize 71} & \underline{\footnotesize 73} \\
    \hline
    Time & {\footnotesize 26} & {\footnotesize 18} & {\footnotesize 21} & {\footnotesize 21} & {\footnotesize 13} & {\footnotesize 17} & {\footnotesize 21} & {\footnotesize 11} & {\footnotesize 16} & {\footnotesize 16} \\
    \hline
    Closure & {\footnotesize 131} & {\footnotesize 37} & {\footnotesize 58} & {\footnotesize 64} & {\footnotesize 29} & {\footnotesize 50} & {\footnotesize 62} & {\footnotesize 25} & {\footnotesize 50} & {\footnotesize 52} \\
    \hline\hline
    \rowcolor{lightgray} Overall & 350 & \textbf{178} & \textbf{232} & 244 & 158 & 227 & \textbf{253} & 143 & 202 & 207 \\
    \hline
    \end{tabular}  
    \label{tab:comparison}
    {\scriptsize Underline indicates the result was evaluated using GPT-3.5-turbo}
\end{table}

\begin{figure}
    \centering
    \includegraphics[width=0.9\linewidth]{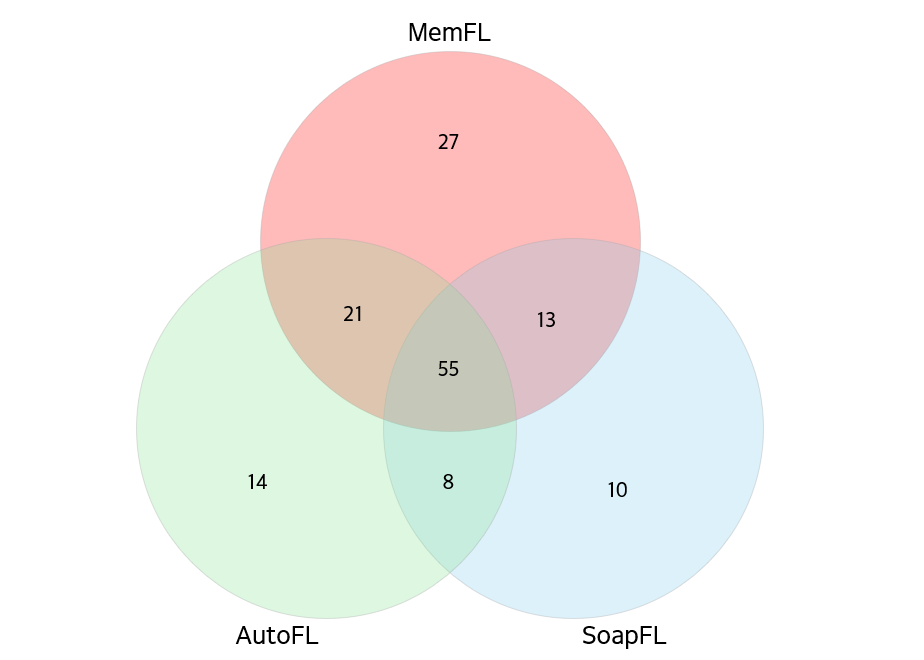}
    \caption{Overlap Analysis}
    \label{fig:enter-label}
\end{figure}

\begin{table}[h]
    \centering
    \renewcommand{\arraystretch}{1.2} 
    \setlength{\tabcolsep}{12pt} 
    \caption{Bug Fixing Cost Comparison}
    \begin{tabular}{|c|c|}
        \hline
        \textbf{Name} & \textbf{Cost per bug (dollars)} \\
        \hline
        SoapFL & 0.0851 \\
        \hline
        AutoFL & 0.0099 \\
        \hline
        MemFL(GPT-4.1-mini) & 0.0094 \\
        \hline
        MemFL(GPT-4o-mini) & 0.0033 \\
        \hline
    \end{tabular}
    \label{tab:cost_comparison}
\end{table}

\textbf{Performance analysis against LLM-base Baselines.}
We first compare MemFL to existing LLM-based methods using GPT-4o-mini, with Dynamic Memory built using the best settings identified in RQ2. As shown in Table 2, MemFL outperforms existing LLM-based baselines in Top-1 and Top-3 accuracy while achieving comparable results in Top-5 accuracy. Specifically, MemFL successfully localized 178 out of 350 bugs at Top-1,  19 more than AutoFL and 35 more than SoapFL, enhancing the top 1 accuracy by 12.7\% and 24.5\% each. The performance improvement is particularly pronounced in the Closure project, where MemFL demonstrates an outstanding 27.6\% and 48\% increase in Top-1 accuracy compared to baselines.

\textbf{Overlap Analysis.}
Figure 5 illustrates the overlap in correctly localized bugs among MemFL, AutoFL, and SoapFL on four projects evaluated using GPT-4o-mini. MemFL uniquely localized 27 bugs, substantially more than either baseline. This demonstrates that MemFL not only covers a large portion of the bugs detected by existing methods but also provides significant additional value by identifying unique faults.

\textbf{Cost and Time analysis against LLM-Based Baselines.}
As shown in Figure 6 and Table 3, MemFL was able to localize faults in an average of 17.4 seconds at the cost of \$0.0033 per bug using GPT-4o-mini. This represents only 21\% of the execution time and 33\% of the cost compared to AutoFL (82.4 seconds, \$0.0099), and just 24\% of the time and 3.9\% of the cost compared to SoapFL (69.9 seconds, \$0.0851). These results highlight MemFL's cost-effectiveness against other LLM-based baselines. Furthermore, cost and time analysis of MemFL with the more recent GPT-4.1-mini (24.71 seconds, \$0.0094\$) reinforces its efficiency on different LLMs.

\begin{figure}
    \centering
    \includegraphics[width=0.95\linewidth]{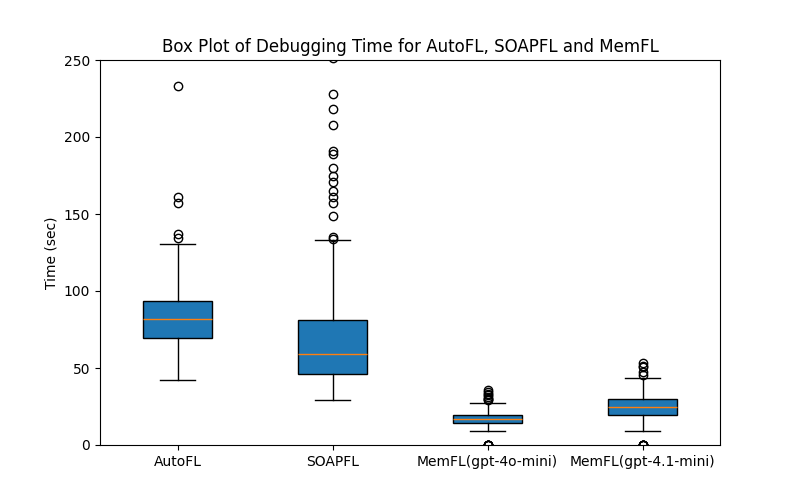}
    \caption{Debugging Time comparison}
    \label{fig:enter-label}
\end{figure}

\begin{table}[h]
    \caption{FL Result Comparison with other baselines}
    \centering
    \footnotesize
    \renewcommand{\arraystretch}{1.5} 
    \setlength{\tabcolsep}{2pt}      
    \definecolor{lightgray}{gray}{0.9}
    \begin{tabular}{l l||ccccc|c}
    \hline
    \textbf{Model} & \textbf{Metric}
      & \textbf{Chart} & \textbf{Lang} & \textbf{Math}
      & \textbf{Time}  & \textbf{Closure}
      & \cellcolor{lightgray}\textbf{Overall} \\
    \hline\hline
    \multicolumn{2}{c||}{\# Bugs}
      & 26 & 61  & 106 & 26  & 131
      & \cellcolor{lightgray}350 \\
    \hline
    \multirow{3}{*}{\shortstack[c]{MemFL\\\scriptsize(GPT-4.1-mini)}}
      & Top1 & 20 & 52 & 68 & 16 & 58  & \cellcolor{lightgray}\textbf{\underline{214}} \\
      & Top3 & 21 & 60 & 86 & 20 & 73  & \cellcolor{lightgray}\textbf{\underline{260}}        \\
      & Top5 & 22 & 60 & 89 & 20 & 82  & \cellcolor{lightgray}\textbf{\underline{273}}       \\
    \hline
    \multirow{3}{*}{\shortstack[c]{MemFL\\\scriptsize(GPT-4o-mini)}}
      & Top1 & 14 & 47 & 62 & 18 & 37  & \cellcolor{lightgray}\textbf{178} \\
      & Top3 & 19 & 57 & 77 & 21 & 58  & \cellcolor{lightgray}\underline{232}          \\
      & Top5 & 21 & 58 & 80 & 21 & 64  & \cellcolor{lightgray}244          \\
    \hline
    \multirow{3}{*}{GRACE}
      & Top1 & 14 & 40 & 61 & 11 & 46  & \cellcolor{lightgray}\underline{172}          \\
      & Top3 & 20 & 51 & 78 & 14 & 69  & \cellcolor{lightgray}\underline{232}          \\
      & Top5 & 22 & 54 & 89 & 19 & 80  & \cellcolor{lightgray}\underline{264}          \\
    \hline
    \multirow{3}{*}{DeepFL$_{\text{cov}}^*$}
      & Top1 & 12 & 43 & 39 &  9 & 64  & \cellcolor{lightgray}167          \\
      & Top3 & 18 & 53 & 68 & 16 & 86  & \cellcolor{lightgray}\textbf{241} \\
      & Top5 & 21 & 56 & 80 & 18 & 97  & \cellcolor{lightgray}\textbf{272} \\
    \hline
    \multirow{3}{*}{Ochiai}
      & Top1 & 6 & 24 & 23 &  6 & 14  & \cellcolor{lightgray}73          \\
      & Top3 & 14 & 44 & 52 & 11 & 30  & \cellcolor{lightgray}151 \\
      & Top5 & 15 & 50 & 62 & 13 &  38 & \cellcolor{lightgray}178\\
    \hline
    \end{tabular}
    \label{tab:transposed_grouped}
    \vspace{4pt}
    \hspace*{8pt}\parbox{\linewidth}{
        \footnotesize
          • \textbf{\underline{Bold and underlined}} indicates the best result. \\
          • \textbf{Bold} indicates the second-best result. \\
          • \underline{Underlined} indicates the third-best result. \\
          • $^*$Indicates that the model was tested on seven additional deprecated data.
    }
\end{table}

\textbf{Performance analysis against other Baselines.}
We also compared MemFL using GPT-4.1-mini and GPT-4o-mini against LBFL and SBFL baselines and found it achieved better overall performance. As shown in Table 4, MemFL with GPT-4.1-mini significantly outperformed baselines, with just 24.71 seconds and \$0.0094 per bug. MemFL with GPT-4o-mini also achieved higher Top-1 accuracy and showed competitive Top-3 and Top-5 accuracy compared to baselines. These results demonstrate that MemFL is effective and efficient, achieving strong results without relying on costly defect datasets or fine-tuning.

\begin{tcolorbox}[colback=gray!20, colframe=white]
\textbf{Answer to RQ 1}: MemFL outperforms existing baselines. Notably, it achieves this while reducing execution time by up to 79\% and cost by 67\%.\end{tcolorbox}

\subsection{RQ2: Optimal Dynamic Memory generation policy}

Various configurations were evaluated to investigate the optimal policy for dynamic memory generation by adjusting the batch size (1, 2, 5, and 10) and the number of iterations (1, 2, and 3), as shown in Figure 7.  The results indicate that MemFL's performance generally converges quickly after one or two iterations. Additionally, while smaller batch sizes exhibit some instability, larger batch sizes (5 and 10) demonstrate more stable performance. Based on performance and computational cost, a batch size of 5 with three iterations is the optimal configuration.

This result can be attributed to the nature of dynamic memory, which retains debugging insights accumulated at each intermediate step of the fault localization process. These insights often take the form of concise, general-purpose guidance that supports effective debugging across different contexts. Such guidance can typically be acquired within one or two iterations and with only a few data points. This characteristic enables a cost-effective and straightforward approach to extracting debugging insights while achieving modest performance improvements.

\begin{figure}
    \centering
    \begin{tikzpicture}
        \begin{axis}[
            width=0.9\columnwidth, 
            height = 6cm,
            xlabel={Dynamic Memory generation Iteration},
            ylabel={Top 1 Accuracy},
            xtick={1,2,3},
            xticklabels={1,2,3},
            ymin=156, ymax=180,
            legend pos=south east,
            legend style={font=\scriptsize},
            grid=major
        ]

        \addplot[dotted, color=blue, mark=o, thick] coordinates {(1,165) (2,167) (3,167)};
        \addlegendentry{Batch size 1}

        \addplot[dash dot, color=green, mark=square, thick] coordinates {(1,172) (2,175) (3,176)};
        \addlegendentry{Batch size 2}

        \addplot[solid, color=red, thick, mark=triangle] coordinates {(1,170) (2,175) (3,178)};
        \addlegendentry{Batch size 5}

        \addplot[dashed, color=purple, mark=diamond, thick] coordinates {(1,171) (2,174) (3,174)};
        \addlegendentry{Batch size 10}

        \end{axis}
    \end{tikzpicture}
    \caption{Top@1 Accuracy over Dynamic memory generation iterations and batch size}
\end{figure}
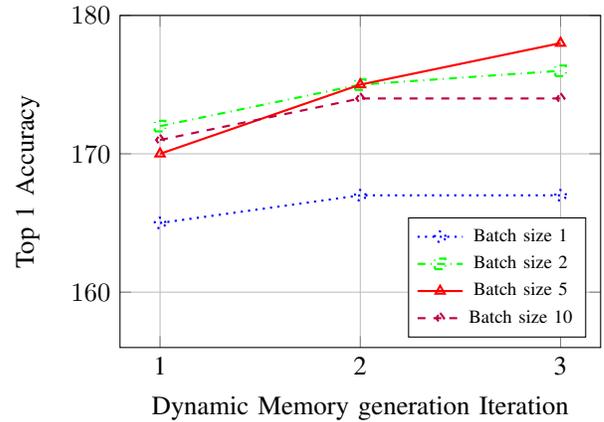

\begin{tcolorbox}[colback=gray!20, colframe=white]
\textbf{Answer to RQ 2}: MemFL’s performance stabilizes after just 1–2 iterations, with larger batch sizes offering greater stability. The optimal setting (batch size 5, three iterations) balances the performance and cost of MemFL \end{tcolorbox}

\subsection{RQ3: Ablation study}

\begin{table}[h]
    \centering
    \caption{Result of ablation study.}
    \label{tab:ablation}
    \begin{tabular}{lccc}
        \toprule
        \textbf{Techniques} & \textbf{Top1} & \textbf{Top3} & \textbf{Top5} \\
        \midrule
        w/o Dynamic Memory & 164 & 229 & 236 \\
        w/o Bug Review Generation   & 162 & 223 & 235 \\
        w/o Code Condensation    & 136 & 176 & 184 \\
        \rowcolor[gray]{0.9} \textbf{MemFL} & \textbf{178} & \textbf{232} & \textbf{244} \\
        \bottomrule
    \end{tabular}
\end{table}

As shown in Table 5, the ablation study demonstrates that all three core components—Dynamic Memory Generation, Bug Review Generation, and Code Condensation—contribute to MemFL's performance. Code Condensation has the most significant impact, followed by Bug Review Generation and Dynamic memory generation. This is because the Code Condensation step directly reduces the search space by hierarchically filtering down to the most relevant classes and methods. 

\begin{tcolorbox}[colback=gray!20, colframe=white]
\textbf{Answer to RQ 3}: All three components enhance MemFL, with Code Condensation having the greatest impact, followed by Bug Review Generation and Debugging Guidance Extraction. \end{tcolorbox}

\section{Discussion}

\subsection{Evaluating the Impact of 5-Fold Cross-Validation}

\begin{table}[h]
    \centering
    \caption{Impact of 5-Fold Cross-Validation}
    \label{tab:ablation}
    \begin{tabular}{cccc}
        \toprule
        \textbf{Techniques} & \textbf{Top1} & \textbf{Top3} & \textbf{Top5} \\
        \midrule
        w/o 5-Fold Cross Validation & 175 & 230 & \textbf{245 }\\
        \rowcolor[gray]{0.9} \textbf{With 5-Fold Cross Validation} & \textbf{178} & \textbf{232} & 244 \\
        \bottomrule
    \end{tabular}
\end{table}

We evaluated MemFL using 5-fold Cross-Validation to prevent potential information leakage during the dynamic memory generation step and compared its performance with and without 5-fold Cross-Validation to assess its impact. As shown in Table 6, the results are similar, suggesting that data leakage had minimal effect. This is likely because the dynamic memory generated in this step is often generic rather than bug-specific, reducing the risk of unintended data leakage.

\subsection{Threats to Validity}

\subsubsection{Internal Validity}
A main internal threat to validity is potential data leakage. The LLMs used in this study, GPT-4o-mini-2024-07-18 and GPT-4.1-mini-2025-04-14, were trained on data up to or beyond October 2023. While the defects4j dataset was released before that, it is possible that some of its content was included in the model's training data. We mitigated this threat by not explicitly naming the dataset (defects4j), but we cannot completely rule out the possibility of indirect exposure.

\subsubsection{External Validity}
MemFL was tested exclusively on Java projects from Defects4J and has not been evaluated on other software ecosystems. However, the dataset consists of five distinct Java projects with over 350 diverse real-world bugs, covering a range of domains and bug types. This diversity helps support the generalizability of our results. However, the effectiveness of its project-specific knowledge approach in different ecosystems remains an open question.

\section{Conclusion}
We propose MemFL, a novel LLM-based fault localization technique that incorporates project-specific knowledge through external memory-based knowledge editing. Existing LLM-based FL approaches rely solely on the general capabilities of the LLM and do not effectively leverage project-specific knowledge. Consequently, they rely on complex processes and exhibit lower performance on more complex projects. In contrast, MemFL introduces a lightweight three-step fault localization process, utilizing a two-component external memory to deliver and iteratively refine project context. This memory structure allows the model to quickly adapt to project intricacies without fine-tuning, significantly enhancing performance.

Evaluations on the Defects4J benchmark demonstrate that MemFL outperforms state-of-the-art LLM-based FL techniques by 12.7 \%  while significantly saving the cost. This efficiency is especially notable in complex projects where existing LLM-based techniques encounter difficulties. Furthermore, we demonstrate that while MemFL achieves comparable performance to LBFL baselines when using GPT-4o-mini, it outperforms all existing baselines with the more recent GPT-4.1-mini. We also evaluated different dynamic memory generation policies. We found that a relatively small dataset combined with a limited number of iterations (specifically, five data points and three iterations) effectively improves fault localization performance. Additionally, our ablation study highlighted the effectiveness of each component of MemFL.

Overall, MemFL provides an accurate, efficient, and practical solution for fault localization.

Future work includes extending the external memory structure to other domains of Software Engineering, such as Automated Program Repair and Vulnerability Detection. We believe that the proposed two-component external memory can be applied to other tasks that benefit from domain-specific knowledge. Additionally, we plan to refine and automate dynamic memory's structure and generation policy to incorporate more detailed and relevant information. This aims to enable AI for SE systems to reason more effectively without relying on fine-tuning or labeled data.

\bibliographystyle{IEEEtran}

\bibliography{references}

\end{document}